-------------------
\documentstyle{laa}
\begin{document}

\thesaurus{02.01.2; 02.08.1; 07.19.1}

\title{Did planet formation begin inside persistent
gaseous vortices?}

\author{P. Barge\inst{1} \and J. Sommeria\inst{2} }
\offprints{ P. Barge }
\institute{Laboratoire d'Astronomie Spatiale, B.P.8 -
13376 Marseille
c\'edex 12
, France \and Ecole Normale Sup\'erieure de Lyon,
Laboratoire de Physique,
CNRS,
46 All\'ee d'Italie,
69364 Lyon c\'edex 07, France}


\date{Received ??, Accepted ??, 1994 }
\maketitle

\begin{abstract}
We explore here the idea, reminiscent in some respect of
Von
Weizs\"acker's
(1944) and Alfv\`en's (1976) outmoded cosmogonies, that
long-lived vortices in a turbulent protoplanetary nebula
can capture large
amount of solid particles and initiate the formation of
planets. Some
puzzling
features of the solar  system appear as natural
consequences of our simple
model:
\par
 - The captured mass presents a maximum near Jupiter's
orbit.\
\par
- Outside this optimal orbit, the collected material,
mainly composed of
low
density particles, sinks deeply into the vortices and
rapidly collapses
into
massive bodies at the origin of the solid core of the
giant planets.\
\par
- Inside this orbit, by contrast, the high density
particles are
preferentially selected by the vortices and assembled by
local
gravitational
instabilities into planetesimals, massive enough to be
released by the
vortices
and to grow later, in successive collisions, to form the
terrestrial
planets.
\

\end{abstract}

Planets are thought to be formed from the dust grains
embedded in a gaseous
disk, probably like  observed around most young low-mass
stars: the
Protoplanetary Nebula. It is likely that  this
by-product of the sun formation
was also, for a while, a turbulent accretion-disk.
During this  stage the star
completes its accretion and the disk spreads outward
with the angular momentum
(justifying the repartition of mass and momentum between
sun and planets).  In
the meantime the dust grains are submitted to a
turbulent diffusion (due to the
gas motions) which speeds up their  growth and enables
to explain the chemical
composition of some meteorites (Morfill, 1983). Due to
collisional
fragmentation, this turbulent coagulation stalls for
centimeter-sized particles
in a highly turbulent nebula (Weidenschilling, 1984) and
for meter-sized
particles in a weakly turbulent nebula (Weidenschilling
and Cuzzi, 1993),
while gravitational binding becomes effective only in
the kilometer range.
So, it is
commonly  thought that the solid material decouples from
the gas only after
some turbulence decayed, in a two stage process:
\
\par
(i) settling of the dust grains toward the mid-plane of
the gaseous disk;
\
\par
(ii) gravitational collapse of the resulting layer of
sediment (when dense
enough) into numerous kilometer-sized bodies, the so
called
"planetesimals".
\
\par
Then, as suggested by the cratering of the present
planets, gravitationaly
bounded bodies grow by the  accumulation of
planetesimals in successive
collisions; this stage  of the planet growth is, indeed,
 reproduced by a
number of dynamical models
(Safronov 1969; Barge and Pellat 1991, 1993).
\
\par
However the above scenario faces two major difficulties.
\par
(1) The solid cores of the giant planets must be formed
in less than some
$10^6$
years, in order for the gas to be captured before being
swept away
(Safronov 1969; Strom {\sl{et al.}} 1993) during
the sun's T-Tauri phase; with a reasonable density of
solid material, this
is
difficult to achieve by planetesimal accumulation
(Safronov 1969; Wetherill 1988), especially for the
outermost planets.\
\par
(2) The formation of the planetesimals themselves is not
clearly
understood.
Indeed the gas, which is supported by a radial pressure
gradient, rotates
at
slightly less than the local Keplerian speed. The
resulting velocity
difference
$\Delta V$ between the sediment layer and the overlying
gas induces shear
turbulence that prevents the layer from settling to the
density required
for
gravitational instability (Weidenschilling and Cuzzi
1993; Cuzzi {\sl{et
al.}}
 1993).\
\par
Both difficulties are solved by the present model, in
which the particles,
once settled in the nebula mid-plane, are captured and
concentrated into
long-lived vortices.\
\par
Such vortices may be maintained by specific instability
mechanism
(Dubrulle 1993), but
more generally emerge from random turbulence in rotating
shear flows.
While
three-dimensional eddies are quickly damped by energy
cascade toward small
scales, two-dimensional turbulence persists without
energy dissipation,
forming
instead larger and larger vortices until a steady
solitary vortex is
formed. Striking examples are the persistent atmospheric
vortices in the
giant
planets (Ingersoll 1990),
like Jupiter's Great Red Spot. This phenomenon can be
reproduced in laboratory experiments (Antipov {\sl{et
al.}} 1986;
Sommeria {\sl{et al.}}
1988; Nezlin and Snezkhin 1993),
and explained in terms of
statistical mechanics of two-dimensional turbulence
(Sommeria {\sl{et al.}} 1991; Miller {\sl{et al.}} 1992;
Michel and Roberts
1994). Observations of
accretion-disks around black-holes (Abramowicz {\sl{et
al.}} 1992)
or T-tauri stars could also indicate
the presence of such organized vortices.
\par
In this letter, as we focus on particle trajectories,
fluid dynamics will
be
discussed at an heuristic level, sufficient to justify a
simple vortex
model.
Particle motions are referred to a cartesian frame, in
which $x$ and  $y$
stand
for the azimuthal position and the radial displacement,
respectively,
rotating
around the sun at the Keplerian angular velocity
$\Omega=\Omega_0 r^{-3/2}$
($\Omega_0$ is the Earth's velocity for $r=1$, in
astronomical unit, AU).
The $y$
axis is directed outward, and the $x$ axis along the
orbital motion, which
has then a clockwise (negative) rotation.
\par
 Under the
standard assumption of hydrostatic balance in the
thickness $H$ of the
nebula ($H\simeq
C_S/\Omega$, where $C_S$ is the sound speed), the
dynamical problem is
two-dimensional.
Further, neglecting pressure forces, the simplest flow
is a set of circular
orbits with
Keplerian azimuthal velocity ($V_x = - 3\Omega y/2$ and
$V_y =0$).
The small
vortices, possibly rising in this flow
with scale $R < H$ and typical vorticity $\Omega$,
have a velocity of the order of $\Omega R$ which is less
than the sound
speed; so, they can be considered as
incompressible.
Then, vortices spinning like the shear flow are robust
and merge
one another.
(while those with opposite sign are laminated by the
shear)
(Marcus 1990; Dowling and Ingersoll 1989).
The process of vortex growth ends up when the Mach
number reaches unity
(i.e)
$R \simeq H$, beyond which energy losses by sound waves
become prohibitive.
Finally, the vortex structure should evolve as to
minimize the pressure
effects
responsible for these losses, the streamlines fitting at
best with the
free-particle trajectories.
\par
An obvious solution is a set of Keplerian ellipses with
the same semi-major
axis but different eccentricities, corresponding in our
rotating frame, to
concentric epicycles ($V_x = - 2\Omega y$, $V_y =\Omega
x/2$); this is a
steady
solution of the fluid equations with uniform pressure. A
correspondence
between
epicyclic motion and vortex flow was used first by Von
Weizs\"acker (1944);
it
appears also in the dynamics of non-axisymmetric
planetary rings where
fluid
streamlines can coincide with particle trajectories and
describe the ring
shapes (Borderies {\sl{et al.}} 1982).
We assume a  simple matching of this "epicyclic flow"
with the
azimuthal Keplerian flow at large distances (see Fig.1):

  $$ \cases{
V_x = - {3\over 2}\Omega y - {1\over 2}\Omega
y~e^{-{{x^2 + y^2}\over {2
R^2}}}
     \cr
V_y = {1\over 2}\Omega x ~e^{-{{x^2 + y^2}\over {2
R^2}}} }~~. $$

\begin{figure}
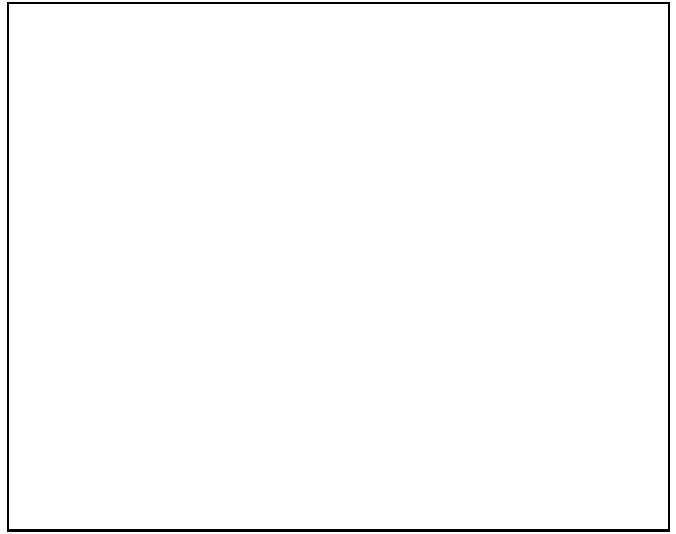

\picplace{7cm}
\caption[]{
Trajectories of the particles captured in a gaseous
vortex, sketched
by the separatrix (dashed line) between open and closed
streamlines. The
particles penetrate into the vortex and spiral inward
toward its center;
they
tend to reach purely epicyclic motion with a transient
behaviour strongly
dependent on the friction parameter: light particles
($\tau_S= 0.05$ in
case (a)) remain near the  edge of the vortex, whereas
heavy ones
($\tau_S= 3$ in case (b)) first sink deeply into the
inner regions. It
must be noted that, for clarity of the figure, the
ordinates have been
expanded
by a factor of two.\
}
\end{figure}
The characteristic size $R$ (or "radius") of this vortex
is limited to the
thickness $H$ of the nebula as discussed above. Its
decay time under persistent
three-dimensional turbulence can be estimated, using the
classical turbulent
viscosity for accretion-disks  $\nu_t = \alpha
C_S^2/\Omega$ with $\alpha =
10^{-3}$ (following current  nebula models
$10^{-4}<\alpha<10^{-2}$). The
corresponding
energy decay time $\tau_D$ is then
estimated by
dividing the
rate of viscous energy dissipation by the vortex kinetic
energy.
This yields $\Omega
\tau_D = 6.5/\alpha$, so that $\tau_D$ is about $500$
rotation periods
$2\pi/\Omega$. For the Great Red Spot of Jupiter or Dark
Oval of Neptune, the
ratio between the estimated friction time ($\sim$ 10
years) and rotation time
(a few days) is similar.
We then similarly expect that, at a given distance from
the sun, successive
mergings have time to produce a unique vortex or at most
very few of them.
By contrast, vortices with sufficient radial separation
(a few times their size
$R\sim H$) cannot merge, so that we finally expect a set
of independent
isolated vortices with radial interval scaling like the
nebula thickness.
Since $H$ scales with a power law of the distance to the
sun
($r^{5/4}$ in the standard model considered below), this
is
consistent with an approximate geometric progression of
the planetary positions.
\par
The particles embedded in the gas of the nebula are
submitted to a friction
drag whose expression depends on the mean-free-path of
the gas molecules
relative to the particle size. For decimetric particles
(and beyond $2$AU from
the sun), mean-free-path exceeds size and the drag
reaches the Epstein regime.
The motion equations of the particles, submitted to the
sun attraction,
Coriolis force and friction drag, then read:

$$ \cases{
{{dv_x}\over {dt}} = - 2\Omega v_y - {1\over {t_S}} (v_x
- V_x)
   \cr
{{dv_y}\over {dt}} = 3\Omega^2 y + 2\Omega v_x - {1\over
{t_S}} (v_y - V_y)
   }~~~, $$
where $ t_S = \rho_d s /(\rho_{gas}C_S)$ is the
stopping-time for a spherical
particle with radius $s$ and density $\rho_d$ in a gas
with density
$\rho_{gas}$. The dynamical evolution depends on the
single non-dimensional
friction parameter $\tau_S =\Omega t_S$, that is on the
particle mass/area
ratio:
(i) the ligthest particles ($\tau_S << 1$) come at rest
rapidly with  the gas
and travel with the local flow; (ii) the heaviest
particles ($\tau_S >> 1$)
cross the vortex with a keplerian motion nearly
unaffected by the friction
drag.
\par
In the intermediate range of $\tau_s$, a numerical
integration of the equations
shows that a particle can be captured by the vortex if
its impact parameter
(initial distance to the $x$ axis) is sufficiently small
(Fig.1); otherwise it
is dragged by the flow.  The corresponding critical
impact parameter $\eta_c$
can be fitted by the function (see Fig.2)

$$ f(\tau_S) = {{\eta_c}\over R} =
{}~{{A~\tau_S^{1/2}}\over {\tau_S^{3/2} +
B}}
   ~~, $$
where $A\simeq 2.4$ and $B\simeq 2.2$. This function
reaches a maximum when
$\tau_S\simeq 1$, and reduces  to the power laws
$\tau_S^{1/2}$ and
$\tau_S^{-1}$, in the limits of the light and heavy
particles, respectively. As
the approach velocity ($3\eta\Omega /2$)  only depends
on the impact parameter,
the mass capture rate is straightforwardly:

$$ {{dM_{capt}}\over {dt}} = {3\over 2} \sigma R^2
\Omega f^2({\tau_S}) $$

where $\sigma$ is the mean surface density of nebular
solid material.
\par

\begin{figure}
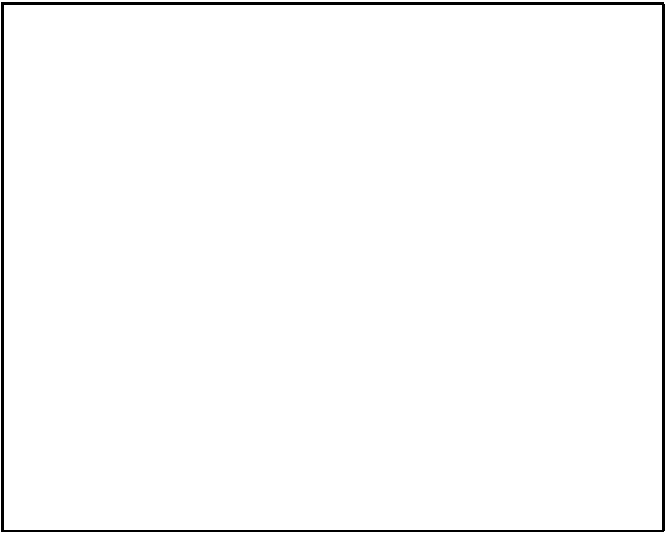

\picplace{7cm}
\caption{
Non dimensional capture cross-section of the vortex as a
function of the
friction parameter ($\tau_S = \Omega t_S$). Filled
squares represent the values
obtained by successive numerical integrations. The
dashed line is the function
$f(\tau_S) = A\tau_S^{1/2}/(\tau_S^{3/2} + B)$ fitting
at best the dependance
$\eta_c(\tau_S)/R$. The curve distinguishes between two
different types of
trajectories: below it, they wind up round the vortex
center and correspond to
a capture; whereas above it, they leave the vortex zone
and correspond to
simple crossing.\
}
\end{figure}

We now evaluate how the capture rate depends on the
distance from the sun,
choosing a standard model of nebula (Cuzzi {\sl{et al.}}
1993)
in which the surface densities
(both
for gas and for particles) and the temperature are the
decreasing
power-laws
$r^{-3/2}$ and $r^{-1/2}$, respectively; at $1$ AU  the
densities are
set to $1 700 gcm^{-2}$ for the gas and to  $20
gcm^{-2}$ for the
particles,
whereas the temperature is assumed to be  $280~K$.
Consequently, the
thickness
of the nebula $H$ ($\simeq R$), approximately $0.04~AU$
near Earth's orbit,
increases as  $r^{5/4}$.
\par
Further, as to get the essence of our capture mechanism,
it is sufficient
to
assume that all the particles have the same density
$\rho_d=2 gcm^{-3}$
(the
density of a composite rock-ice material)  and the same
size $s_* = 40cm$
(a
typical prediction in a  gravitationally stable layer of
sediment
(Weidenschilling and Cuzzi 1993),
inside
which fragmentation is  less effective than in a fully
turbulent
accretion-disk).
The friction parameter, which writes $\tau_S = 2~{\rho_d
s/\sigma_{gas}}$,
increases as $r^{3/2}$.  As a result the capture rate,
proportional to
$f^2(\tau_S)$, is optimum at the distance $r_*$ from the
sun for which the
function $rf^2$ is maximum. With our numerical values
this optimum is
reached
when $r_* \simeq 7.5~AU$, that is in between the
present Jupiter's and
Saturn's orbits, explaining the predominance of these
two planets.
The mass collected (at constant rate)  reaches typical
planetary values
after a
time $\Delta t$ corresponding to $500$ revolutions of
the vortex (see table
I).
\
\begin{table}
\caption[]{Amount of captured mass}
\begin{flushleft}
\begin{tabular}{lllll}
\hline
 $r~(A.U.)$ &  $\Delta t~(yrs)$ &  $\tau_S$ &
$M_{capt}(M_\oplus)$ &
$M_{core}(M_\oplus)~^a$ \\
\hline
$~1$ &  $5.00~10^2$ &  $~0.09$ &  $~0.6$ &  $  -  $ \\
$~2$ &  $1.41~10^3$ &  $~0.27$ &  $~3.2$ &  $  -  $ \\
$~5$ &  $6.00~10^3$ &  $~1.05$ &  $16.0$ &  $15-30$ \\
$10$ &  $1.45~10^4$ &  $~2.97$ &  $18.0$ &  $16-23$ \\
$20$ &  $4.20~10^4$ &  $~8.42$ &  $~7.8$ &  $11-13$ \\
$30$ &  $8.25~10^4$ &  $15.46$ &  $~3.8$ &  $14-16$ \\
\hline
\end{tabular}
\end{flushleft}
{Note: }
(a) Classical estimation of the amount of high-Z
material contained in the
giant planets (Pollack 1985);
notice that these values are still under  debate and, as
recently proposed (Guillot {\sl{et al.}} 1994), could be
significantly
smaller.\
\end{table}
\
The decrease of the predicted masses at large distance
seems a bit too
strong,
when compared to the estimated masses (Pollack 1985)
of heavy elements contained in the
giant planets; in fact, it would be reduced by
accounting for a dispersion
in
particle size and density. On the other hand, comparison
with the masses of
the
terrestrial  planets has been discarded as requiring the
further modelling
of
collisional accumulation.\par

The above calculations implicitly assume that the
particles are
continuously
renewed near the vortex orbit. This occurs due to the
inward drift under
the
systematic drag associated with the velocity difference
$\Delta V$ between
gas
and particles.
This drift, indeed, which reaches its optimum value
$\Delta V$ for $\tau_S \sim 1$ (that is  near Jupiter's
orbit), results in
a
mass  flux exceeding easily the capture rate (We have
also introduced this
drift
in the expression for the gas velocity $V_x$ and check
that it has no
influence on the
capture cross section).
 \par This two dimensional
capture mechanism adds to the vertical settling toward
the nebula midplane,
which is
known to form  a particle sublayer whose typical
thickness is  $H_p\simeq
10^{-3} H$
(Cuzzi {\sl{et al.}} 1993). It results, inside the
vortices, in an
increasing  surface
density $\sigma_{vort}$ and in a stronger
volume density $\sigma_{vort}/H_p$ which reaches much
more easily the
Roche threshold for gravitational instability. In terms
of the particle
velocity dispersions $C_p\sim \Omega H_p$ the criterion
for instability to
occur reads:\
\par

$$ C_p \le {{\pi G \sigma_{vort}}\over {\Omega} }~~~, $$
where $G$ is the gravitational constant.  In the absence
of any surface
density
enhancement ($\sigma_{vort} = \sigma$), this velocity
threshold is very
low,
$20 cms^{-1}$ in our nebula model, and is easily
exceeded by any residual
turbulence.  Indeed, according to the classical model of
turbulent
accretion-disk, the  velocity dispersion $C_p = \alpha
C_S$ (with $\alpha =
10^{-3}$) is of the  order of $2 ms^{-1}$ at 1 AU. Even
with an initially
laminar nebula, a minimal velocity dispersion
(Weidenschilling and Cuzzi 1993; Cuzzi {\sl{et al.}}
1993)
$C_p = 2\Delta V/Re^*$
(where $R_* \sim 100$) would be generated by the
turbulent shear between
the
particle sublayer (considered  as heavy fluid) and the
overlying gas. This
velocity difference $\Delta V$  between gas and
particles is approximately
$60ms^{-1}$ at any distance from the sun and the
resulting velocity
dispersion
(several $ms^{-1}$) is sufficient  to inhibit
gravitational
instability.\
\par
By contrast, inside the vortices,  the surface density
is increased by
several
order of magnitude in some ten  rotation periods, so
that gravitational
instabilities become much easier and rapidly gather the
material into
planetesimals.\
\par

The  fate of these planetesimals (nearly insensitive to
the gas friction) will, in fact, strongly depends on the
friction parameter
$\tau_S$ of the particles they  formed from, that is  on
the distance from
the
sun. Indeed preliminary computations indicate that:\
\par
(1) Inside Jupiter's orbit ($\tau < 1$), particle
concentration occurs in
an
annular region at the vortex periphery, as clearly seen
from the dotted
trajectory of Fig.1; the resulting planetesimals have
wide epicyclic
oscillations and are quickly released from the vortex.
Afterward the growth
 of
the terrestrial planets would proceed following the
"standard"
collisional history (Safronov 1969; Barge and Pellat
1991, 1993).\
\par
(2) Outside Jupiter's orbit ($\tau > 1$), the trapped
particles deeply sink
into the central vortex region (solid trajectory of
Fig.1) and reach an
epicyclic, nearly ballistic motion;  once formed,
planetesimals
remain along similar slowly-evolving orbits, until they
collapse
into a single body, massive enough to form a Giant
Planet after the
capture of the surrounding gas (Safronov 1969; Pollack
1985).\
\par
The direction of planetary rotation depends on the
subsequent phases of dust
contraction and gas accretion, and the result is far
from obvious (Dones and
Tremaine, 1993; Coradini et al. 1989). However, it is
straigthforward to show
that the angular momentum of a swarm of particles
contracting under
self-gravity and inelastic collisions, when refered to
its center of mass, is
conserved, in the standard inertial frame of reference;
then, simple
calculations indicate that this angular momentum is
prograde like with
Keplerian circular orbits (whereas vorticity remains
retrograde, even in the
inertial frame).
\par
 In summary, after decay of the initial
three-dimensional turbulence and
settling of the solid  particles toward the nebula
mid-plane, two-dimensional
turbulence could persist for a long time and organizes
into long-lived vortices
able to strongly concentrate the solid material. This
allows us to suggest new
solutions to some major problems in the modelling of
planetary formation:\
\par
(i) the gravitational instability at the origin of the
planetesimals is made
easier,\
\par
(ii) the cores of the four giant planets form in less
than $10^5~yrs$, while
the terrestrial planets result from longer planetesimal
accumulation.\
\par
Another important consequence of our model is simply
related to the fact that,
in a given vortex, particles with $\tau_S\sim 1$ are
preferentially captured.
This corresponds to dense particles inside the optimal
radius $r^*$ and to
light ones outside. An efficient mechanism of chemical
segregation is therefore
provided by the mass/area dependance of the friction
parameter; it could help
explaining some of the strong disparities observed in
the compositions of
meteorites and planets.\
\par
Of course the existence and structure of our long-lived
vortices would require
further dynamical justification.
However it must be stressed that the capture mechanism
we describe is
unsensitive
to the choice of the starting assumptions (i.e) nebula
model and constant
particle size. Indeed, all the conclusions of the paper
hold as long as the
friction time $\tau_S$ increases with the sun distance,
reaching unity at
Jupiter's orbit. This is due to the fact that $\tau_S$,
which depends on the
ratio $\Omega/(\rho_{gas} C_S)$, strongly increases with
the distance from the
sun, a property which holds for various nebula
structures and a wide
range of particle sizes, and vortex shapes (provided its
core fits with
epicyclic motion).
Our model has therefore a strong predictive
potentiality, and it is reasonable
to consider it as a possible and fruitful alternative to
the classical scenario
of planetesimal formation.\

\acknowledgements{This work was part of the GDR program
"Dynamique des Fluides
G\'eophysiques et Astrophysiques".}

\end{document}